\documentclass[10pt, conference, compsocconf]{IEEEtran}
\usepackage{cite}
\usepackage{balance}
\usepackage{color}
\usepackage{epsfig}
\usepackage{epstopdf}
\usepackage{graphicx}
\usepackage{tabularx}
\usepackage{booktabs}

\usepackage{varwidth}

\usepackage{amssymb,amsmath,amsthm}

\usepackage{bm}
\usepackage{algorithm}%

\newfloat{routine}{htbp}{loa}
\floatname{routine}{Routine}
\makeatletter\newcommand\l@subroutine{\@dottedtocline{1}{1.5em}{2.3em}}\makeatother

\usepackage{algpseudocode}
\usepackage{pict2e}
\makeatletter
\def\BState{\State\hskip-\ALG@thistlm}

\makeatother

\newcommand{\m}[1]{\mathcal{#1}}

\usepackage{colortbl}

\newcounter{remarkCounter}

\newcounter{probCounter}

\begin{document}

\title{Collaborative Multi-bitrate Video Caching and Processing in Mobile-Edge Computing Networks}


\author{\IEEEauthorblockN{Tuyen X. Tran, Parul Pandey, Abolfazl Hajisami, and Dario Pompili}
\IEEEauthorblockA{Department of Electrical and Computer Engineering\\
Rutgers University--New Brunswick, NJ, USA\\
E-mails: \{tuyen.tran, parul$\_$pandey, hajisamik, pompili\}@cac.rutgers.edu}
}

\maketitle

\thispagestyle{empty}

\begin{abstract}
Recently, Mobile-Edge Computing (MEC) has arisen as an emerging paradigm that extends cloud-computing capabilities  to the edge of the Radio Access Network (RAN) by deploying MEC servers right at the Base Stations (BSs). In this paper, we envision a collaborative joint caching and processing strategy for on-demand video streaming in MEC networks. Our design aims at enhancing the widely used Adaptive BitRate (ABR) streaming technology, where multiple bitrate versions of a video can be delivered so as to adapt to the heterogeneity of user capabilities and the varying of network condition. The proposed strategy faces two main challenges: (i) not only the videos but their appropriate bitrate versions have to be effectively selected to store in the caches, and (ii) the transcoding relationships among different versions need to be taken into account to effectively utilize the processing capacity at the MEC servers. To this end, we formulate the collaborative joint caching and processing problem as an Integer Linear Program (ILP) that minimizes the backhaul network cost, subject to the cache storage and processing capacity constraints. Due to the NP-completeness of the problem and the impractical overheads of the existing offline approaches, we propose a novel online algorithm that makes cache placement and video scheduling decisions upon the arrival of each new request. Extensive simulations results demonstrate the significant performance improvement of the proposed strategy over traditional approaches in terms of cache hit ratio increase, backhaul traffic and initial access delay reduction. 

\end{abstract}

\section{Introduction}
{\bf{Motivation:}} Over the last few years, the proliferation of Over-The-Top (OTT) video content providers (YouTube, Amazon Prime, Netflix,...), coupled with the ever-advancing multimedia processing capabilities on mobile devices, have become the major driving factors for the explosion of on-demand mobile video streaming. According to the prediction of mobile data traffic by Cisco, mobile video streaming will account for 72$\%$ of the overall mobile data traffic by 2019~\cite{cisco2019global}. While such demands create immense pressure on mobile network operators, distributed edge caching has been recognized as a promising solution to bring video contents closer to the users, reduce data traffic going through the backhaul links and the time required for content delivery, as well as help in smoothing the traffic during peak hours. In wireless edge caching, highly sought-after videos are cached in the cellular Base Stations (BSs) or wireless access points so that demands from users to the same content can be accommodated easily without duplicate transmissions from remote servers. 

Recently, Mobile-Edge Computing (MEC)~\cite{tran2017commag, hu2015mobile, zhang2016energy, ahmed2016survey, liu2016delay, mao2016dynamic} has been introduced as an emerging paradigm that enables a capillary distribution of cloud computing capabilities to the edge of the cellular Radio Access Network (RAN). In particular, the MEC servers are implemented directly at the BSs using generic-computing platforms, enabling context-aware services and applications in close-proximity to the mobile users. With this position, MEC presents an unique opportunity to not only implement edge caching but also to perform edge processing. In this paper, we aim at exploiting  MEC storage and processing capabilities to improve caching performance and efficiency beyond what could be achieved using traditional approaches. 

Due to the heterogeneity of users' processing capabilities and the variation of network condition, user preference and demand towards a specific video might be different. For example, users with highly capable devices and fast network connection usually prefer high resolution videos while users with low processing capability or low-bandwidth connection may not enjoy high quality videos because the delay is large and the video may not fit within the device's display. Leveraging such behavior, Adaptive Bit Rate (ABR) streaming techniques~\cite{stockhammer2011dynamic, akhshabi2011experimental} have been widely used to improve the quality of delivered video on the Internet as well as wireless networks. In ABR streaming, the quality (bitrate) of the streaming video is adjusted according to the user device's capabilities, network connection, and specific request. Existing video caching systems often treat each user request equally and independently, whereby each bitrate version of a video is offered as a disjoint stream (data file) to the user, which is a waste of storage.

{\bf{Our vision:}} \emph{In contrast to most of the existing works on video caching which are not ABR-aware and mainly rely on the ``store and transmit'' mechanism without any processing, our work proposes to utilize both caching and processing capabilities at the MEC servers to satisfy users' requests for videos with different bitrates. To the best of our knowledge, we are the first to introduce collaborative joint caching and processing in MEC networks.} Specifically, owing to its real-time computing capability, a MEC servers can perform transcoding of a video to different variants to satisfy the user requests. Each variant is a bitrate version of the video and a higher bitrate version can be transcoded to a lower bitrate version. For example, a video at bit-rate of 5 Mbps (720p) can be transcoded from the same video at bit-rate of 8 Mbps (1080p). Moreover, \emph{we extend the collaborative caching paradigm to a new dimension where MEC servers can assist each other to not only provide the requested video via backhaul links but also transcode it to the desired bitrate version} (for example, when the requesting server's processing load is full). In this way, the requested variant of a video can be transcoded by any MEC server on the delivery path from where the original video is located (data provider node) to the home MEC server (delivery node) of the end user. The potential benefits of this strategy is three-fold: (i) the original remote content server does not need to generate different bitrate versions of the same video, (ii) users can receive videos that are suited for their network condition and multimedia processing capabilities as content adaptation is more appropriately done at the network edge, and (iii) collaboration among the MEC servers enhances cache hit ratio and balance processing load in the network.

{\bf{Challenges and contributions:}} The proposed strategy, however, faces several challenges. Firstly, caching multiple bitrate versions of the videos incurs high overhead in terms of storage. Although hard disk is very cheap nowadays, it is neither cost-efficient nor feasible to store all these files. Secondly, real-time video transcoding is a computation-intensive task. Transcoding of a large number of videos simultaneously might quickly exhaust the available processing resource on the MEC servers. Therefore, it is very important to design a caching and request scheduling scheme that efficiently utilizes both the given cache and processing resouces. To this end, we formulate the collaborative joint caching and processing problem as an Integer Linear Program (ILP) that minimizes the backhaul network cost, subject to the cache storage and processing capacity constraints. Due to the NP-completeness of the problem and the impractical overheads of the existing offline approaches, we adopt the popular Least Recently Used (LRU) caching policy and propose a novel online video scheduling algorithm that makes decision upon arrival of each new request. It should be noted that our approach does not need \emph{a-priori} information about the content popularity and request arrivals as commonly assumed.

{\bf{Related Works:}} In general, content caching has been extensively studied in the context of Information Centric Network (ICN) (see for example~\cite{fricker2012impact, xie2012enhancing}  and the references therein). In~\cite{hajimirsadeghi2015joint, hajimirsadeghi2016joint}, the authors develop game theoretic models to evaluate joint caching and pricing strategies among access networks, transit networks and content providers in an ICN. 
Different from the ICN settings, considerable research efforts have focused on content caching in wireless networks \cite{bastug2014living, golrezaei2012femtocaching, ahlehagh2014video}, and on exploiting the backhaul links connecting the BSs for collaborative caching~\cite{Gharaibeh2015online, wang2014cache}. Recently, the authors in \cite{tran2016octopus, tran2016cooperative} proposes a cooperative hierarchical caching in a Cloud Radio Access Network (C-RAN) where the cloud-cache is introduced as a bridging layer between the edge-based and core-based caching schemes. The authors propose a low complexity, online cache management strategy, consisting of a proactive cache distribution algorithm and a reactive cache replacement algorithm, to minimize the average delay cost of all content requests. Along this line, work in~\cite{Mosleh2016Globecom} proposes a coordinated data assignment algorithm to minimize the network cost with respect to both the precoding matrix and the cache placement matrix in a C-RAN. However, the heterogeneity of networks and user capabilities have not been considered in these works to facilitate  ABR video streaming. 


To account for multi-bitrate video streaming, a number of works have focused on Scalable Video Coding (SVC)~\cite{zhu2013design, poularakis2016caching, yu2016enhancing}. However, SVC is not preferred in industry in the past, which is partly due to the lack of hardware decoding support, and especially it may significantly increase power consumption on mobile devices whose battery capacity is limited. 

The works in \cite{shen2004caching, pedersen2016enhancing} consider caching and processing for muli-bitrate (or multi-version) video streaming, which are closest to our work. However they only study on one cache entity, as opposed to the collaborative scheme of multiple caching/processing servers in our paper. Furthermore, the proposed technique in \cite{pedersen2016enhancing} resolves the optimization problem from scratch every time there is a new request arrival, thus resulting in re-directing large numbers of pre-scheduled requests. On the other hand, the heuristic solution in \cite{zhao2015version} requires the knowledge of the content popularities, which may be hard to estimate accurately in practice.

{\bf{Paper organization:}} The remainder of this paper is organized as follows: In Section~\ref{sec:model}, we describe considered caching system. In Section~\ref{sec:online_alg}, we formulate the joint collaborative caching and processing problem and present the proposed online algorithm. Section~\ref{sec:results} presents our simulation results. Finally we conclude the paper in Section~\ref{sec:conclusion}.


\section{MEC Caching System} \label{sec:model}
In this section, we present the envisioned distributed caching system deployed on MEC networks, followed by the settings of the considered model.

\subsection{System Architecture}
\begin{figure}
 \centering
 \includegraphics[width=0.45\textwidth]{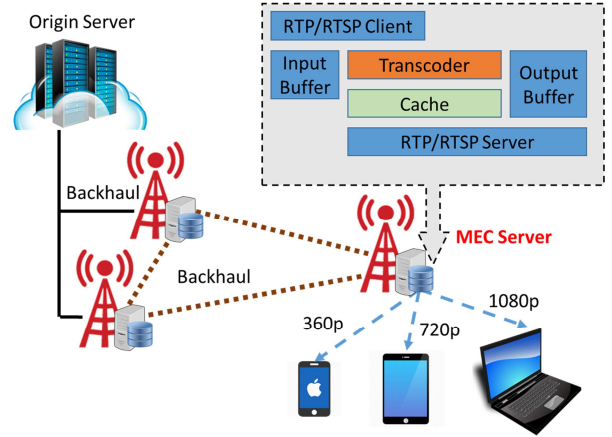}
\caption{Illustration of collaborative video caching and processing framework deployed on MEC network. The cache server implemented on MEC server acts as both RTP/RTSP client and server.}\label{fig:mec_cache}
\end{figure}

\begin{figure*}[t]
 \centering
\includegraphics[scale = .8]{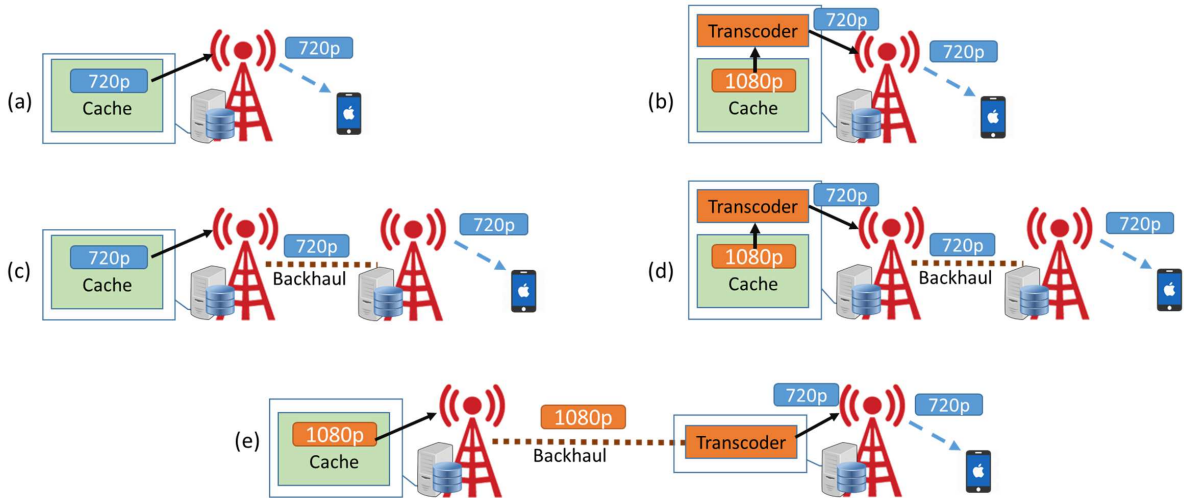}
\caption{Illustration of possible (exclusive) events that happen when a user request for a video. (a) The video is obtained from cache of the home BS; (b) a higher bitrate version of the video from cache of the home BS is transrated to the desired bitrate version and deliver to the user; (c) the video is retrieved from cache of a neighboring BS or from the origin content server; (d) a higher bitrate version of the video from cache of a neighboring BS is transrated using the co-located transcoder and is then transfered to the home BS; (e) similar to (d) but the transcoding is done at the home BS's transcoder.}\label{fig:cases}
\end{figure*}

As shown in Fig.~\ref{fig:mec_cache}, a MEC network consists of multiple MEC servers connected via backhaul links. Each MEC server is deployed side-by-side with the BS in a cellular RAN, providing computation, storage and networking capabilities to support context-aware and delay-sensitive applications in close proximity to the users. In this paper, we envisage the use of MEC servers for enabling video caching and processing. The concept of MEC cache server is similar to the cache proxy server in the Internet~\cite{shen2004caching}, however we consider these servers in a collaborating pool that could share content and processing resources. In particular, each cache server acts as a client to the origin content server (in the Internet) and to other peer cache servers. An RTP/RTSP client is built into the server to receive the streamed content from other servers via backhaul links and put it into the input buffer. If needed, the transcoder will transcode the input stream to a desired bitrate stream and pushes it out to the output buffer; otherwise the input buffer is directly moved to the cache and/or output buffer for transmitting to the end users. Here, an RTP/RTSP server is built to stream the video to the end users and to other servers. The data in the output buffer is obtained either from the transcoder or from the cache. In Fig.~\ref{fig:cases}, we illustrated the possible (exclusive) events that happen when a user request for a video.


Video transcoding, i.e., compressing a higher bitrate video to a lower bitrate version, can be done by various techniques~\cite{vetro2003video}. Among those, compressed domain based approaches, such as bitrate reduction and spatial resolution reduction, are the most favorable~\cite{shen2004caching}. In general, video transcoding is a computation-intensive task. The cost of a transcoding task can be regarded as the CPU usage on the MEC cache server.

\subsection{Settings}
In this paper, we consider a MEC network of $K$ cache servers, denoted as $\m{K} = \left\{ {1,2,...K} \right\}$. Each cache server is attached to a BS in the cellular RAN that spans $K$ cells. Additionally, $k = 0$ denotes the origin content server. The MEC servers are connected to each other via backhaul mesh network. The collection of videos is indexed as $\m{T} = \left\{ {1,2,...V} \right\}$. Without loss of generality, we consider that all videos have the same length and each has $L$ bitrate variants. Hence, the size of each video variant $l$, denoted as $r_l\left[ {{\rm{bytes}}} \right]$, is proportional to its bitrate. The set of all video variants that a user can request is $\m{V} = \left\{ {{v_l}\left| {v \in \m{T},l = 1,...L} \right.} \right\}$. In the subsequent analysis, unless otherwise stated, we will refer to video and video variant interchangeably. We consider that video $v_l$ can be transcoded from video $v_h$ if $l \leq h$ and the cost (CPU usage) of transcoding $v_h$ to $v_l$ is denoted as $\phi_{hl}$, $\forall v \in \m{T}$ and $l,h = 1,...L$. As considered in \cite{pedersen2016enhancing}, we assume that $p_{hl}$ is proportional to $r_l$, i.e., $p_{hl} = p_l = \tau r_l$. It should be noted that this cost model can be easily extended to the case where $p_{hl}$ depends on both $r_h$ and $r_l$.

In this paper, we consider that video requests arriving at each BS following a Poisson process with rate $\lambda_j$, $j\in \m{K}$. The caching design is evaluated in a long time period to accumulate a large number of request arrivals. The set of new request arriving at BS $j$ in the considered time period is denoted as $\m{N}_j \subseteq \m{V}$. 

We consider that each user only connects to and receives data from the nearest BS (in terms of signal strength), which is later referred to as the user's \emph{home BS}. Further extension to the system employing Coordinated Multi-Point transmission (CoMP), where each user can be served by multiple BSs, is a subject for future investigation. In the considered MEC caching system, each cache server is provisioned with a storage capacity of $M_j\left[ {{\rm{bytes}}} \right]$. 
To describe the cache placement, we define the variables $c_j^{vl} \in \left\{ {0,1} \right\},j \in K,{v_l} \in \m{V}$, in which $c_j^{vl} = 1$ if $v_l$ is cached at server $j$ and $c_j^{vl} = 0$ otherwise. The cache storage capacity constraint at each server $j \in \m{K}$ can be expressed as,
\begin{equation} \label{eq:cache_cap}
\sum\limits_{{v_l} \in \m{V}} {{r_l}c_j^{vl}}  \le {M_j}, \forall j \in \m{K}.
\end{equation}

To describe the possible events that happen when a request for video $v_l \in \m{N}_j$ arriving at server $j$, we introduce the binary variables $\left\{ {x_j^{vl},y_j^{vl},z_{jk}^{vl},t_{jk}^{vl},w_{jk}^{vl}} \right\} \in \left\{ {0,1} \right\}$, which are explained as follows.

\begin{itemize}
\item $x_j^{vl} = 1$ indicates that $v_l$ can be served directly from cache of BS $j$, given that $c_j^{vl} = 1$ (as illustrated in Fig.~\ref{fig:cases}(a)); and $x_j^{vl} = 0$ otherwise. 
\item $y_j^{vl} = 1$ when $v_l$ is retrieved from cache at BS $j$ after being transcoded from a higher bitrate variant (as illustrated in Fig.~\ref{fig:cases}(b)); and $y_j^{vl} = 0$ otherwise. 

\item $z_{jk}^{vl} = 1$ if $v_l$ is retrieved from cache of BS $k \neq j, k \in \m{K} \cup \left\{ 0 \right\}$ (including the remote server, as illustrated in Fig.~\ref{fig:cases}(c)); $z_{jk}^{vl} = 0$ otherwise.

\item $t_{jk}^{vl} = 1$ when $v_l$ is obtained by transrating a higher bitrate version from cache of BS $k \neq j, k \in \m{K}$ and the transcoding is performed at BS $k$ (as illustrated in Fig.~\ref{fig:cases}(d)); $t_{jk}^{vl} = 0$ otherwise. 
\item $w_{jk}^{vl} = 1$ when $v_l$ is obtained by transrating a higher bitrate version from cache of BS $k \neq j, k \in \m{K}$ and the transcoding is performed at BS $j$ (as illustrated in Fig.~\ref{fig:cases}(e)); $w_{jk}^{vl} = 0$ otherwise.
\end{itemize}

When a video is requested, it will be served following one of the event described above. To ensure this, we impose the following constraint ($\forall j \in \m{K}, v_l \in \m{V}$),
\begin{equation} \label{eq:con1}
x_j^{vl} + y_j^{vl} + \sum\limits_{k \ne j,k \in \m{K}} {\left( {z_{jk}^{vl} + t_{jk}^{vl} + w_{jk}^{vl}} \right)}  + z_{j0}^{vl} = 1.
\end{equation}

\subsection{Backhaul Network Cost}
Let $d_{jk}$ denote the backhaul cost incurred when the $j$th cache server retrieves a video of unit size from the $k$th cache server, and let $d_{j0}$ denote the backhaul cost incurred when the $j$th cache server retrieves a video of unit size from the origin content server in the Internet. If we associate a cost between any two directly connected BSs, then for any two BSs $j$ and $k$, we can calculate $d_{jk}$ using the minimum cost path between $j$ and $k$. In practice, ${d_{j0}}$ is usually much greater than ${d_{jk}}$ as the backhaul link connecting a BS to the origin content server is of many-fold further than the backhaul links between the BSs. This makes it cost-effective to retrieve content from the in-network caches whenever possible rather than downloading them from the remote server. To reflect this cost model, as considered in~\cite{wang2014cache, Gharaibeh2015online, tran2016octopus}, we set ${d_{j0}} \gg {d_{jk}},\forall j,k \in \m{K}$.

The incurred backhaul cost when serving request for video $v_l$ from BS $j$ can be calculated as ($\forall j \in \m{K}, v_l \in \m{V}$), {\small
\begin{equation}
{D_j}\left( {{v_l}} \right) = {r_l}\left[ {{d_{j0}} z_{j0}^{vl} + \sum\limits_{k \ne j,k \in \m{K}} {{d_{jk}}\left( {z_{jk}^{vl} + t_{jk}^{vl} + w_{jk}^{vl}} \right)} } \right].
\end{equation}}
The backhaul cost reflects the amount of data traffic going through the backhaul links, and thus the resource consumption of the network. On the other hand, reducing the backhaul cost (by retrieving content from shorter paths) also directly translates to the decrease in initial delay that the users have to wait before starting to play the videos. Therefore, it is very important to minimize the backhaul cost of serving video requests, which constitutes a large portion in the total backhaul cost of a cellular network. 

\section{Joint Collaborative Video Caching and Processing} \label{sec:online_alg}
Here we formulate the collaborative joint caching and processing problem and present the offline optimal solution, followed by the proposed online approach.

\subsection{Problem Formulation}
To realize the envisioned joint collaborative caching and processing in a MEC network, we now formulate the optimization problem that aims at minimizing the total backhaul cost of serving all the video requests. In particular, given the available resources (cache storage and processing capability), the objective is to jointly determine (i) a \emph{cache placement policy}, i.e., deciding $\left\{ {c_j^{vl}} \right\}$ and (ii) a \emph{video request scheduling policy}, i.e., deciding $\left\{ {x_j^{vl},y_j^{vl},z_{jk}^{vl},t_{jk}^{vl},w_{jk}^{vl}} \right\}$. The problem formulation is as follows,
\begin{subequations} \label{eq:prob}
\begin{align} \label{eq:prob_a}
\hspace{-0.1cm} &\mathop {\min } \sum\limits_{j \in \m{K}} {\sum\limits_{{v_l} \in {\m{N}_j}} {{D_j}\left( {{v_l}} \right)} } ,\\
\label{eq:prob_b}
\hspace{-0.1cm} \text{s.t.} \hspace{0.1cm} &x_j^{vl} \le c_j^{vl},\hspace{0.5cm} \forall j \in \m{K},{v_l} \in \m{V},\\ \label{eq:prob_c}
&z_{jk}^{vl} \le c_k^{vl},\hspace{0.5cm}\forall j, k \in \m{K},\,{v_l} \in \m{V}, \\
 \label{eq:prob_d}
&y_j^{vl} \le \min \left( {1,\sum\limits_{m = l + 1}^L {c_j^{vm}} } \right), \hspace{0.5cm} \forall j\in \m{K},{v_l} \in \m{V},\\ \label{eq:prob_e}
&t_{jk}^{vl} \le \min \left( {1,\sum\limits_{m = l + 1}^L {c_k^{vm}} } \right),\hspace{0.5cm} \forall j\in \m{K},{v_l} \in \m{V}, \\ \label{eq:prob_f}
&w_{jk}^{vl} \le \min \left( {1,\sum\limits_{m = l + 1}^L {c_k^{vm}} } \right),\hspace{0.5cm} \forall j, k\in \m{K},{v_l} \in \m{V},\\ \nonumber 
&x_j^{vl} + y_j^{vl} + \sum\limits_{k \ne j,k \in \m{K}} {\left( {z_{jk}^{vl} + t_{jk}^{vl} + w_{jk}^{vl}} \right)}  + z_{j0}^{vl} = 1, \\\label{eq:prob_g}
&\hspace{6cm}\forall j \in \m{K}, \\
\label{eq:prob_h} 
&\sum\limits_{{v_l} \in \m{V}} {{r_l}c_j^{vl}}  \le {M_j}, \forall j \in \m{K}, \\
 \nonumber
&\hspace{-0.4cm}\sum\limits_{{v_l} \in {\m{N}_j}} {{p_l}\left( {y_j^{vl} + \sum\limits_{k \ne j,k \in \m{K}} {w_{jk}^{vl}} } \right) + \sum\limits_{k \ne j,k \in \m{K}} {\sum\limits_{{v_l} \in {\m{N}_k}} {{p_l}t_{kj}^{vl}} } }  \le {P_j}, \\   \label{eq:prob_i}
&\hspace{6cm} \forall j \in \m{K}, \\
& c_j^{vl}, {x_j^{vl},y_j^{vl},z_{jk}^{vl},t_{jk}^{vl},w_{jk}^{vl}}  \in \left\{ 0, 1 \right\}, \hspace{0.5cm} \forall j \in \m{K}, v_l \in \m{V}.
\end{align}
\end{subequations}

The constraints in the formulation above can be explained as follows: constraints (\ref{eq:prob_b}) and (\ref{eq:prob_c}) ensure availability of the exact video variants; constraints (\ref{eq:prob_d}), (\ref{eq:prob_e}) and (\ref{eq:prob_f}) ensure the availability of the higher bitrate variants for transcoding; constraint (\ref{eq:prob_g}) ensures that each request should only be fulfilled by one unique path as mentioned in (\ref{eq:con1}); constraint (\ref{eq:prob_h}) ensures the cache storage capacity; finally constraint (\ref{eq:prob_i}) ensures the availability of processing resource (in terms of encoded bits that can be processed per second) for transcoding at each cache server. 

The problem in (\ref{eq:prob}) is an ILP and is NP-complete, which can be shown by reduction from a  multiple knapsack problem~\cite{gary1979computers}. Thus, solving this problem to optimal in polynomial time is extremely challenging. A common approach to make such problem more tractable is to rely on continuous relaxation of the binary variables to obtain fractional solutions (where a video request is served from multiple places and a video can be partially stored in the cache). While the fractional solutions satisfy the constraints, simply rounding them to integer solutions will lead to infeasible solutions. Another approach is to resolve the optimization problem everytime there is a new request arrival; however this will result in re-directing large numbers of pre-scheduled requests and wasting buffer data. Another key challenge of solving problem (\ref{eq:prob}) in practice is that the complete set of request arrivals, i.e., $\m{N}_j$'s, are not known in advance. Furthermore, we make no assumption about the popularity of the contents, and thus $\m{N}_j$'s are not known probabilisticly, either. 

Motivated by the aforementioned drawbacks, we adopt the popularly used Least Recently Used (LRU) cache placement policy~\cite{lee2001lrfu}, and propose a new \emph{online} Joint Collaborative Caching and Processing (JCCP) algorithm that makes cache placement and video request scheduling decisions upon each new arrival of video request. In the following, before presenting our proposed online JCCP algorithm, we briefly discuss its offline counterpart to serve as a performance benchmark. 

\subsection{Offline Approach}
The LRU cache placement policy fetches the video from the neighboring caches or the origin content server upon user request if it is not already cached at the home BS. It then saves the content in the cache and if there is not enough space, the entries that have been least recently used are evicted to free up space for the newly added content. The LRU-based offline approach to problem (\ref{eq:prob}) will recompute the optimal request scheduling everytime there is a new arrival or departure. The \emph{offline} request scheduling problem is expressed as in (\ref{eq:prob1}), where $\m{N}_j^*$ is the set of videos currently being served at BS $j \in \m{K}$.


Note that the solution of the offline problem is optimal in the long run. However such solution might cause re-directing the existing video requests whenever the optimal request scheduling solution is re-calculated, thus wasting the buffered data at the BSs. Another drawback of the offline solution is that the complexity of solving the problem scales with the number of request arrivals and number of caching servers and thus it is highly impractical to re-solve this problem, which is an integer program, when there is a large number of request arrivals in a very short time. 
\begin{subequations} \label{eq:prob1}
\begin{align} \label{eq:prob1_a}
\hspace{-0.1cm} &\mathop {\min } \sum\limits_{j \in \m{K}} {\sum\limits_{{v_l} \in {\m{N}_j^*}} {{D_j}\left( {{v_l}} \right)} } ,\\
\label{eq:prob1_b}
\hspace{-0.1cm} \text{s.t.} \hspace{0.1cm} &x_j^{vl} \le c_j^{vl},\hspace{0.5cm} \forall j \in \m{K},{v_l} \in \m{V},\\ \label{eq:prob1_c}
&z_{jk}^{vl} \le c_k^{vl},\hspace{0.5cm}\forall j, k \in \m{K},\,{v_l} \in \m{V}, \\
 \label{eq:prob1_d}
&y_j^{vl} \le \min \left( {1,\sum\limits_{m = l + 1}^L {c_j^{vm}} } \right), \hspace{0.5cm} \forall j\in \m{K},{v_l} \in \m{V},\\ \label{eq:prob1_e}
&t_{jk}^{vl} \le \min \left( {1,\sum\limits_{m = l + 1}^L {c_k^{vm}} } \right),\hspace{0.5cm} \forall j\in \m{K},{v_l} \in \m{V}, \\ \label{eq:prob1_f}
&w_{jk}^{vl} \le \min \left( {1,\sum\limits_{m = l + 1}^L {c_k^{vm}} } \right),\hspace{0.5cm} \forall j, k\in \m{K},{v_l} \in \m{V},\\ \label{eq:prob1_g}
&x_j^{vl} + y_j^{vl} + \sum\limits_{k \ne j,k \in \m{K}} {\left( {z_{jk}^{vl} + t_{jk}^{vl} + w_{jk}^{vl}} \right)}  + z_{j0}^{vl} = 1, \\
 \nonumber
&\hspace{-0.4cm}\sum\limits_{{v_l} \in {\m{N}_j^*}} {{p_l}\left( {y_j^{vl} + \sum\limits_{k \ne j,k \in \m{K}} {w_{jk}^{vl}} } \right) + \sum\limits_{k \ne j,k \in \m{K}} {\sum\limits_{{v_l} \in {\m{N}_k^*}} {{p_l}t_{kj}^{vl}} } }  \le {P_j}, \\   \label{eq:prob_i}
&\hspace{6cm} \forall j \in \m{K}, \\
& {x_j^{vl},y_j^{vl},z_{jk}^{vl},t_{jk}^{vl},w_{jk}^{vl}}  \in \left\{ 0, 1 \right\}.
\end{align}
\end{subequations}

\subsection{Proposed Online JCCP Algorithm} 
In the following, we present the proposed online algorithm for the joint collaborative caching and processing problem, which bases on the LRU cache replacement policy. The proposed online JCCP algorithm makes video request scheduling decision immediately and irrevocably upon each video request arrival at one of the BSs. 

Denote $\m{N^*} = \left( {{\m{N}_1^*},...{\m{N}_K^*}} \right)$ as the set of videos currently being served in the system, where $\m{N}_j^*$ is served at BS $j$, we can calculate the current processing load (due to transcoding) at BS $j$ as, {\small
\begin{equation}
U_j(\m{N}^*) = \sum\limits_{{v_l} \in {\m{N}_j^*}} {{p_l}\left( {y_j^{vl} + \sum\limits_{k \ne j, k \in \m{K}} {w_{jk}^{vl}} } \right) + \sum\limits_{k \ne j, k \in \m{K}} {\sum\limits_{{v_l} \in {\m{N}_k^*}} {{p_l}t_{kj}^{vl}} } }.
\end{equation}}

We define the \emph{closest} (in terms of bitrate) transcodable version of video $v_l$ at BS $j$ as $T\left( {j,{v_l}} \right) = {v_h}$, in which,
\begin{equation}
h = \mathop {\arg \min }\limits_{m > l} c_j^{vm}\,\,\,\,\,\,{\rm{s}}{\rm{.t}}{\rm{.}}\,\,\,\,\,{\rm{c}}_j^{vm} = 1.
\end{equation}

For each video request $v_l$ arriving at BS $j\in \m{K}$, we present the cache placement and request scheduling decisions made by the online JCCP algorithm as in Algorithm~\ref{alg:LRU}. In particular, the algorithm starts with empty cache at each BS and new video fetched to each cache will be updated following the LRU policy. For each new request for $v_l$ at BS $j$, if $v_l$ cannot be directly retrieved (step 2) or transcoded (step 3) from cache of BS $j$, the algorithm will search for $v_l$ or its transcodable version from other neighboring caches. Step 6 finds the exactly requested video $v_l$ from the neighboring caches, and if that exists, $v_l$ will be retrieved from the cache with lowest backhaul cost. Otherwise, a transcodable version of $v_l$ will be searched from neighboring caches in step 7. If the transcodable version exists in the cache of BS $k$, the algorithm will select the cache server (either server $k$ or the requesting server $j$) with most available processing resource to perform transcoding. Finally, if $v_l$ cannot be satisfied by the cache system, it will be fetched from the origin content server (in step 18), which incurs the highest backhaul cost. 

\begin{algorithm} 
\caption{Online JCCP}\label{alg:LRU}
\renewcommand{\Statex}{\item[\hphantom{\bfseries Step \arabic{ALG@line}.}]}
\begin{algorithmic}[1]
\State Initialize: $c_j^{vl} = 0, \forall v_l \in \m{V}, j\in \m{K}$

\State For each video request $v_l$ arriving at BS $j\in \m{K}$, proceed.

\If{$c_j^{vl} = 1$} stream $v_l$ from cache of BS $j$ to the user. 
\ElsIf{$T\left( {j,{v_l}} \right) \ne \emptyset $ and ${U_j}\left( {{\m{N}^*}} \right) + {p_l} \le {P_j}$}
		
	\State \begin{varwidth}[t]{0.9\linewidth}
    transcode $T\left( {j,{v_l}} \right)$ from cache of BS $j$ to $v_l$ and then stream it to the end user.
    \end{varwidth}
    \vspace{0.1cm}	 
	
\ElsIf{$\sum\limits_{k \ne j,k \in \m{K}} {c_k^{vl}}  \ge 1$}
	\State $f = \mathop {\arg \min }\limits_{k \ne j, k \in \m{K}} d_{jk}\,\,\,\,\,\,{\rm{s}}{\rm{.t}}{\rm{.}}\,\,\,\,\,c_k^{vl} = 1$
	
	\State \begin{varwidth}[t]{0.9\linewidth}
    retrieve $v_l$ from cache of BS $f$ to BS $j$ and then stream it to the end user.
    \end{varwidth}
    \vspace{0.1cm}	
    
\ElsIf{$\bigcup\limits_{k \ne j,k \in \m{K}} {T\left( {k,{v_l}} \right)}  \ne \emptyset $}

	\State Calculate ${Q_k}\left( {{\m{N}^*}} \right) = {P_k} - {U_k}\left( {{\m{N}^*}} \right) - p_l,\forall k \in \m{K}$.
	\State $f = \mathop {\arg \max }\limits_{k \ne j,k \in {\rm{ }}\m{K}} {Q_k}\left( {{\m{N}^*}} \right).$
	
	\If{${Q_f}\left( {{\m{N}^*}} \right) \geq 0$}
		\State transcode $T\left( {f,{v_l}} \right)$
 to $v_l$ at cache of BS $f$.
 		\State \begin{varwidth}[t]{0.8\linewidth}
    		retrieve $v_l$ from cache of BS $f$ to BS $j$ 	and then stream it to the end user.
    		\end{varwidth}
    		\vspace{0.1cm}	
 
 	\Else{ continue.}
		
	\EndIf
	
\Else
	\State \begin{varwidth}[t]{0.9\linewidth}
    retrieve $v_l$ from the origin content server and then stream it to the end user.
    \end{varwidth}
    \vspace{0.1cm}	

\EndIf

\State Update $c_j^{vl}$, $\forall j \in \m{K}, v_l \in \m{V}$ following LRU policy.

\end{algorithmic}
\end{algorithm}

\section{Performance Evaluation} \label{sec:results}
In this section, we evaluate the performance of the proposed joint collaborative caching and processing solution under various cache sizes, processing capacities and video request arrival rates. 
We consider a MEC networks consisting of 3 MEC servers, each deployed on a BS of a cellular RAN. We assume the video library $\m{V}$ that consists of $V=1000$ unique videos, each having $4$ bitrate variants. Like in~\cite{pedersen2016enhancing}, we set the relative bitrates of the four variants to be $0.82, 0.67, 0.55$ and $0.45$ of the original video bitrate ($2~\rm{Mbps}$). We assume that all video variants have equal length of $10~\rm{minutes}$. The popularity of the videos being requested at each BS follows a Zipf distribution with the skew parameter $\alpha = 0.8$, i.e, the probability that an incoming request is for the $i$-th most popular video is given as,
\begin{equation}
{q_i} = \frac{{1/{i^\alpha }}}{{\sum\nolimits_{j = 1}^V {1/{j^\alpha }} }}.
\end{equation}
In order to obtain a scenario where the same video can have different popularities at different locations, we randomly shuffle the distributions at different BSs. For each request, one of the four variants of the video is selected with equal probability. Video requests arrive one-by-one at each BS $j$ following a Poisson distribution with rate ${\lambda _j}\left[ {\rm{reqs}/\rm{min}} \right]$. For each simulation, we randomly generate $10,000$ requests at each BS. The end-to-end latency of fetching video content from the local BS, from a neighboring BS, and from the origin content server are randomly assigned following the uniform distribution in the ranges $[5,10](\rm{ms})$, $[20,50](\rm{ms})$, and $[100,200](\rm{ms})$, respectively~\cite{li2015delay}. The backhaul cost $d_{j0}$'s and $d_{jk}$'s are set equal to the corresponding delays. In terms of resources, we set the cache storage capacity relative to the total size of the video library, and the processing capacity is regard as the number of encoded bits that can be processed per second.

In our performance evaluation, we consider the following three important metrics: \emph{(i) cache hit ratio} - the fraction of requests that can be satisfied either by retrieving from the cache or by transcoding; \emph{(ii) average access delay} $[\rm{ms}]$ - average latency of the contents travelling from the caches or the origin server to the requesting user; \emph{(iii) external backhaul traffic load $[\rm{TB}]$} - the volumn of data traffic going through the backhaul network due to users downloading videos from the origin server. 

In the simulation results, we refer to our proposed joint collaborative caching and processing scheme as \emph{Online-JCCP}. We compare the performance of \emph{Online-JCCP} with the \emph{Offline-Optimal} solution as described in Section~III-B and two baselines described below.
\begin{itemize}
\item \emph{CachePro}: A joint caching and processing scheme  without collaboration among the cache servers, as proposed in~\cite{pedersen2016enhancing}.
\item \emph{CoCache}: A collaborative caching scheme without transcoding, and the LRU cache placement policy is employed. 
\end{itemize} 

\subsection{Impact of cache size and processing capacities}

\begin{figure*}[t]
 \centering
 \begin{tabular}{ccc}
\hspace*{-.3cm}\includegraphics[scale = .6]{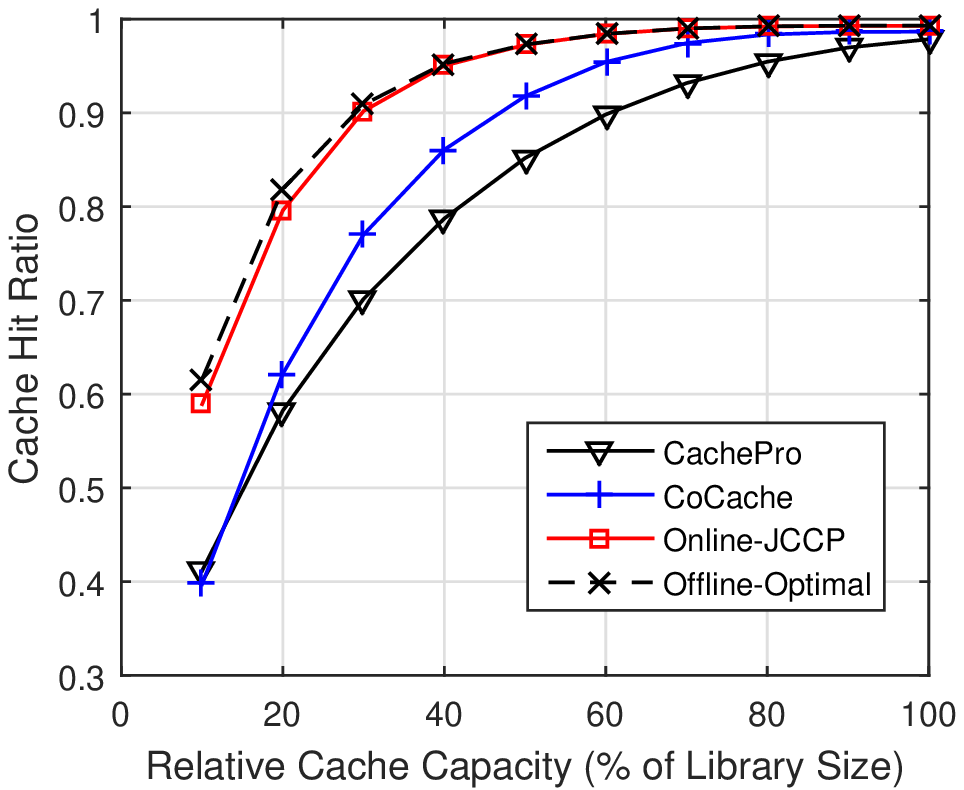} &
\hspace*{-.6cm}\includegraphics[scale = .6]{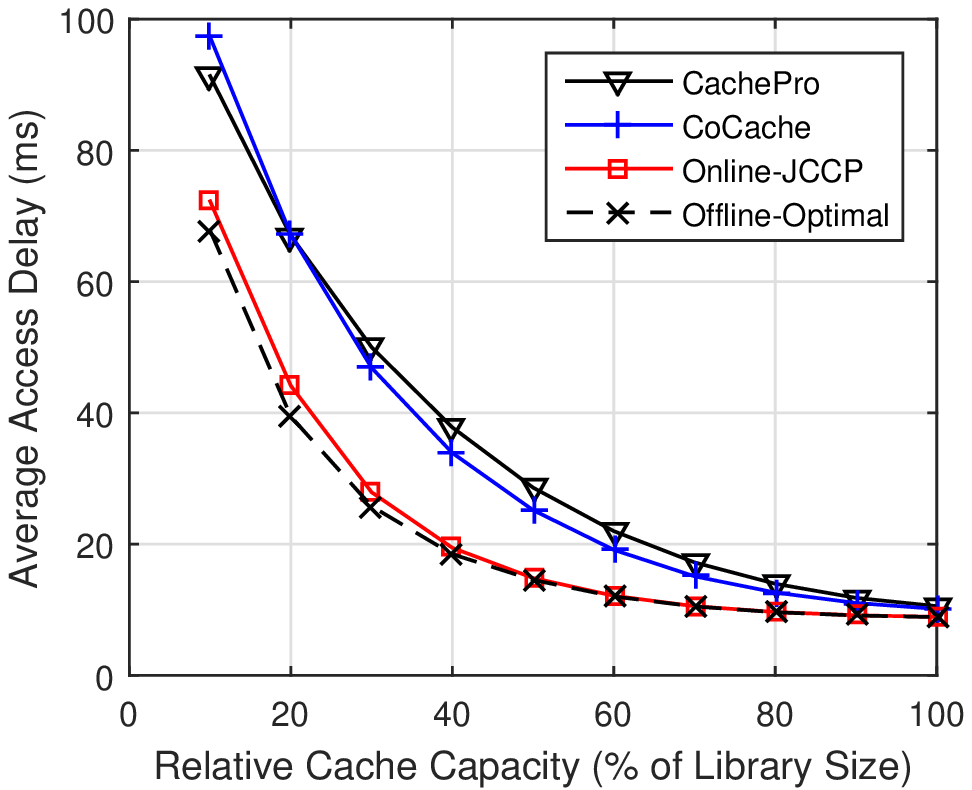} &
\hspace*{-.6cm}\includegraphics[scale = .6]{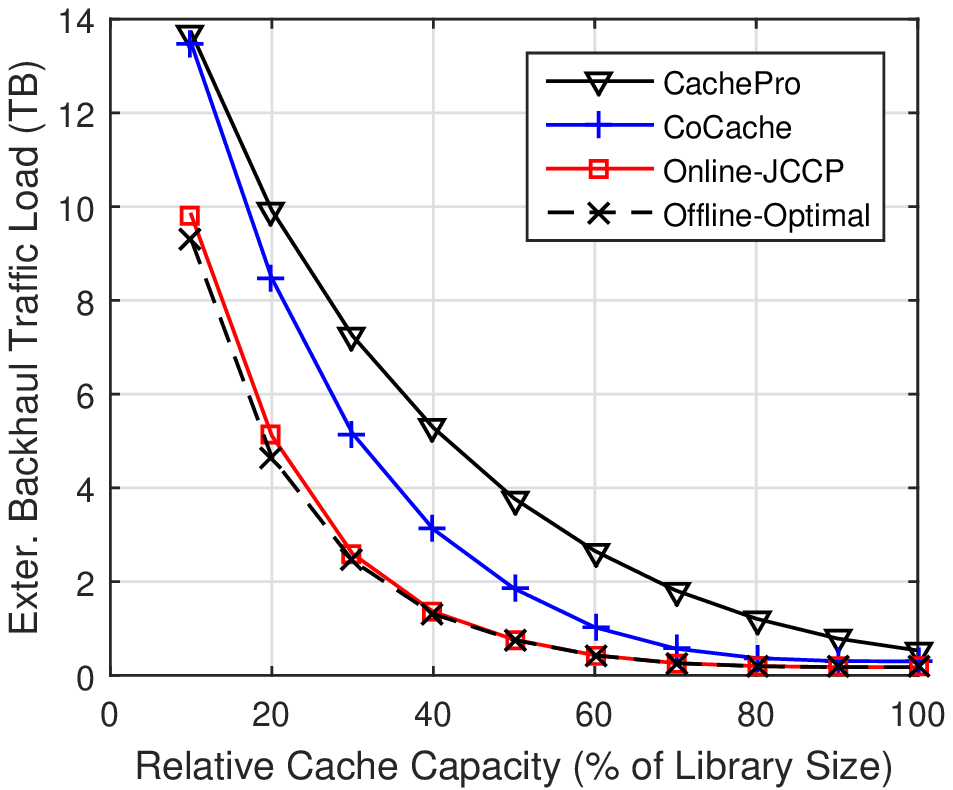} \\
 \small(a) & \small(b) & \small(c)
\end{tabular}
\caption{Performance comparison of different caching schemes when increasing relative cache capacity at each server; $P_j = 10~\rm{Mbps}, \lambda_j = 8~reqs/minute, \forall j \in \m{K}$.
}\label{fig:vsCacheCap}
\end{figure*}

\begin{figure*}[t]
 \centering
 \begin{tabular}{ccc}
\hspace*{-.3cm}\includegraphics[scale = .6]{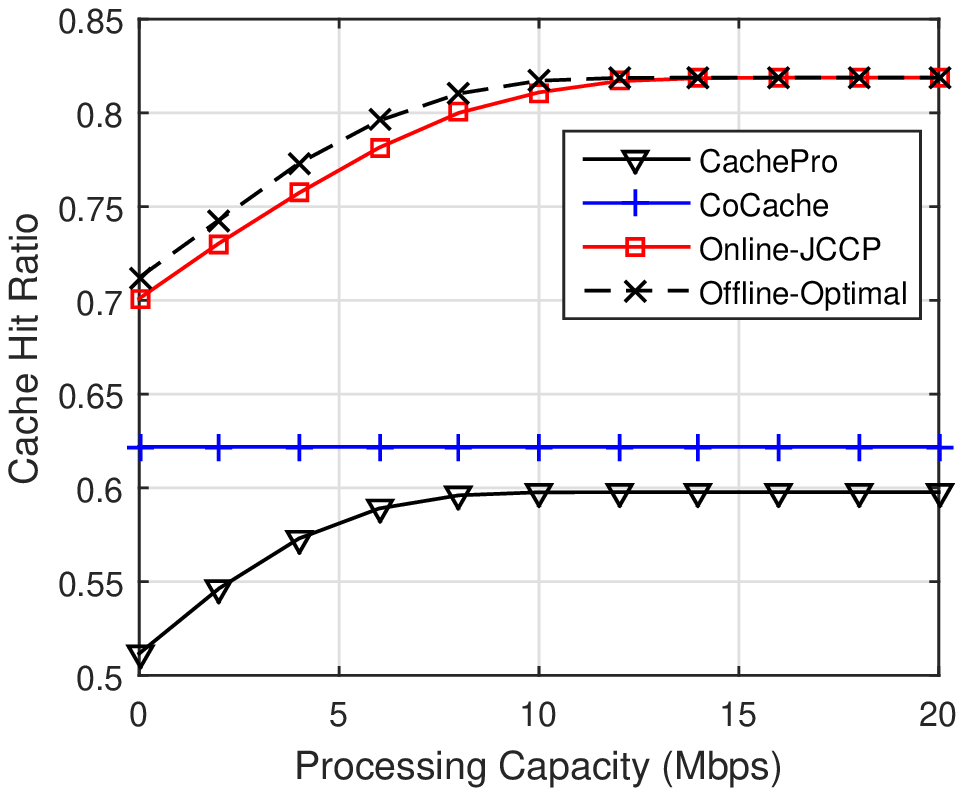} &
\hspace*{-.6cm}\includegraphics[scale = .6]{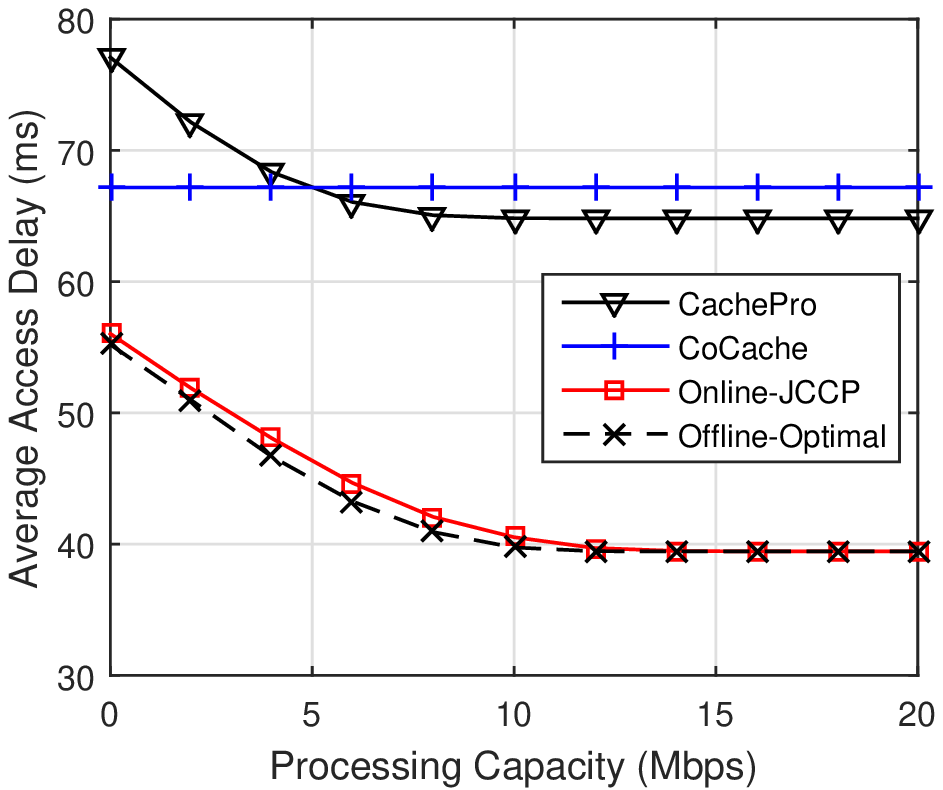} &
\hspace*{-.6cm}\includegraphics[scale = .6]{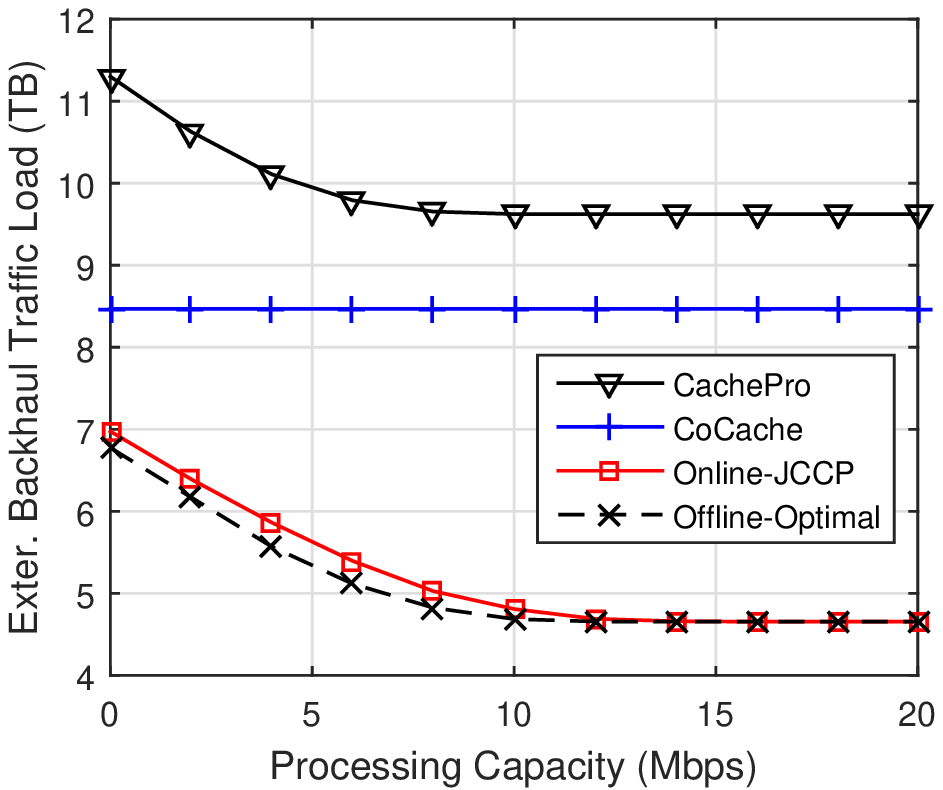} \\
 \small(a) & \small(b) & \small(c)
\end{tabular}
\caption{ Performance comparison of different caching schemes when increasing relative cache capacity at each server; $M_j = 20\% [\rm{Library~Size}], \lambda_j = 8~ reqs/minute, \forall j \in \m{K}$.
}\label{fig:vsProCap}
\end{figure*}

We compare the performance of the four considered caching schemes in terms of cache hit ratio, average access delay and external backhaul traffic load at different relative cache sizes as in Fig.~\ref{fig:vsCacheCap}(a, b, c) and at different processing capacities as in Fig.~\ref{fig:vsProCap}(a, b, c). From the figures, we can see that increasing cache size and processing capacity always result in performance improvement in all schemes. Notice that the \emph{Online-JCCP} scheme significantly outperforms the two baselines at a wide range of cache and processing capacities. At moderate cache and processing capacities, the performance of \emph{Online-JCCP} scheme is slightly lower than that of the optimal scheme; however when the cache size and processing capacity are high, the performance of \emph{Online-JCCP} is the same as that of the optimal scheme. Notice from Fig.~\ref{fig:vsProCap} that the performance improvement diminishes at certain processing capacity, from which the performance of \emph{Online-JCCP} and \emph{Offline-Optimal} schemes are almost identical.

\subsection{Impact of request arrival rate}
\begin{figure}
 \centering
 \includegraphics[width=0.45\textwidth]{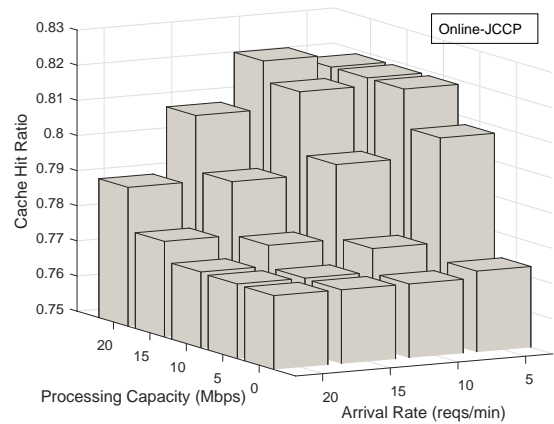}
\caption{Hit ratio performance of the Online-JCCP algorithm at different values of video request arrival rate and processing capacity; $M_j = 20\% [\rm{Library~Size}], \forall j \in \m{K}$.}\label{fig:3Dhit}
\end{figure}

In Fig.~\ref{fig:3Dhit}, we illustrate the cache hit ratio performance of the \emph{Online-JCCP} scheme at different values of video request arrival rate and processing capacity. It can be seen that the cache hit ratio decreases at high request arrival rates and low processing capacity, and it increases otherwise.  

\begin{figure}
 \centering
 \includegraphics[width=0.45\textwidth]{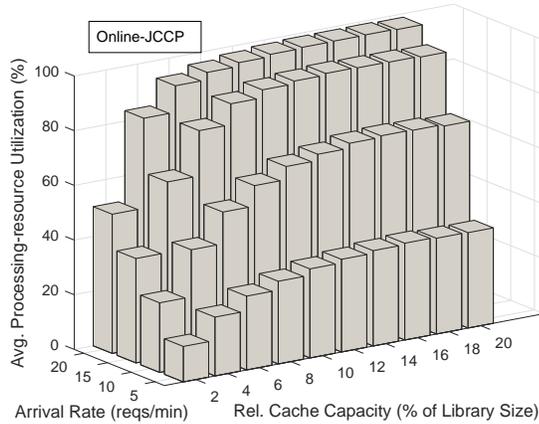}
\caption{Average processing resource utilization at the cache servers using Online-JCCP algorithm; $P_j = 20~\rm{Mbps}, \forall j \in \m{K}$.}\label{fig:3Dutil}
\end{figure}

Fig.~\ref{fig:3Dutil} illustrates the processing resource utilization of \emph{Online-JCCP} scheme versus different video request arrival rates and cache capacities. We observe that the processing utilization increases with arrival rate and moderate cache capacity, however it decreases at high cache capacity. This can be explained as when the cache capacity is high, the MEC servers can store a large number of video variants and thus there are fewer requests requiring transcoding.

\section{Conclusions} \label{sec:conclusion}
In this paper, we propose the idea of deploying a collaborative caching in a multi-cell Mobile-Edge Computing (MEC) networks, whereby the MEC servers attached to the BSs can assist each other for both caching and transcoding of multi-bitrate videos. The problem of joint collaborative caching and processing is formulated as an Integer Linear Program (ILP) aiming at minimizing the total cost of retrieving video contents over backhaul links. Due to the NP-completeness of the problem and the absence of the request arrival information in practice, we proposed an efficient online algorithm, referred to as JCCP, that makes cache placement and video request scheduling decisions upon arrival of each new request. Extensive simulation results have demonstrated the significant performance improvement of the proposed JCCP scheme in terms of cache hit ratio, content access delay, and external backhaul traffic load, over the traditional approaches. Furthermore, while the performance of JCCP is slightly lower than that of the offline optimal scheme at moderate cache storage and processing capacities, the performance gap is approaching zero when the caching and processing resources are high. 

\textbf{Acknowledgment: }This work was supported in part by the National Science Foundation Grant No. CNS-1319945.


\bibliographystyle{ieeetr}\small
\bibliography{wons16_short}


\end{document}